\documentclass[12pt]{article}
\usepackage{graphicx}
\usepackage{url}
\usepackage{pstricks}
\usepackage{color}
\usepackage{epsfig}
\usepackage{fancyhdr}
\usepackage{mathbbol}
\usepackage{pstricks}
\usepackage{color}
\usepackage{bbm}
\usepackage{dsfont}
\usepackage{multirow}
\usepackage{hyperref}
\usepackage{bm}
\usepackage{amsmath}
\usepackage{amsfonts}
\usepackage{amssymb}

\def\be{\begin{equation}}
\def\ee{\end{equation}}
\def\bea{\begin{eqnarray}}
\def\eea{\end{eqnarray}}

\title{A $J^{PG}=1^{++}$ Charged Resonance in  the $Y(4260)\to \pi^+\pi^- J/\psi$ Decay?}
\author{L. Maiani$^{\P}$$^*$, V. Riquer$^{\P}$\\
R. Faccini$^*$, F. Piccinini$^\dag$, A. Pilloni$^*$, A.D. Polosa$^*$\\
\small{$^{\P}$ CERN-PH-TH, Geneva 1211, Switzerland}\\
\small{$^*$Dipartimento di Fisica and INFN, Sapienza Universit\`a di Roma}\\
\small{ Piazzale Aldo Moro 2, I-00185 Roma, Italy}\\
\small{$^\dag$INFN Sezione di Pavia, Via Bassi 6, I-27100 Pavia, Italy} 
}
\date{\today}

\begin{document}
\maketitle
\begin{abstract}
\noindent 
New BES and Belle data show a peak in the $Y(4260)$ decay into $J/\psi$ plus one charged pion. We point out that the peak might correspond to a charged resonance at about  $3880$~MeV predicted time ago within a tetraquark model. The same tetraquark model predicts another peak at about 100~MeV below the  observed one. We discuss the possibility of having it in present data. On the other hand we expect that if the molecular picture were the correct one, a peak corresponding to a $D^*\bar D^*$ state should appear at  about $4020$~MeV. 
\end{abstract}

\pagenumbering{arabic}

\section*{Introduction}
Resonances with hidden charm or beauty, denoted by $X$ and $Y$, not fitting the simple quarkonium state, have been discovered by Belle and Babar~\cite{revs} and have opened a new field in hadron spectroscopy. The case has been reinforced by the discovery of charged, hidden charm or beauty states, the so-called $Z$ states~\cite{z4430,zb}, which evidently require two quarks and two antiquarks, going beyond the meson, quark-antiquark, paradigm.

The BESIII collaboration reported in~\cite{bes} the discovery of a charged hidden charm axial meson, $Z_c^+$, with a mass of $3899\pm 6$~MeV and a width of \mbox{$\Gamma=46\pm 22$~MeV}
decaying into $J/\psi\,\pi^+$. A week later, the Belle collaboration confirmed the discovery~\cite{belle}, reporting a mass of $3895\pm 8$~MeV and a width $\Gamma=63\pm35$~MeV. Some evidence of the $Z_c^+$ and of the neutral partner $Z_c^0$ has also been found in CLEO data~\cite{cleo} in the $\psi(4160)$ decay. The simplest quantum numbers assignment is $J^{PG}=1^{++}$,  $G$ being the \mbox{$G$-parity}.
Such a particle is predicted in the tetraquark scheme of Ref.~\cite{Maiani:2004vq} at a mass of $3882$~MeV, very close to the position of the observed peak. In the tetraquark scheme, the  particle should be accompanied by a lower mass one, $(Z^\prime_c)^+$, with the same quantum numbers and a predicted mass of $3755$~MeV.

We point out that in the molecular scheme~\cite{mol} one expects, analogously, three axial mesons one being the $X(3872)$, the second at about the mass of the peak we are considering but the third one around the $D^*{\bar D^*}$ threshold, which is some $100$~MeV above the present peak. A precise determination of the mass distribution of  $\pi^\pm J/\psi$ in the decay $Y(4260)\to \pi^+\pi^- J/\psi$ could lead to a discrimination between the two alternatives. 
In this paper such hypotheses are tested by combining the BES and Belle final invariant mass spectra. 

We also attempt an estimate of decay rates, which, within large uncertainties, seem to support the rather large observed width. In this context, we stress the importance of a measurement of the neutral channels, $Y(4260)\to \pi^0\pi^0 \,J/\psi$ or $Y(4260)\to \pi^0\eta \,J/\psi$, to identify the neutral partner of  $Z_c^+(1^{++})$, which is expected to decay into $J/\psi\; \pi^0$ and possibly  in $J/\psi \;\eta$ in presence of sizable isospin violation.

Finallty we note that two $Z_b$ resonances with hidden beauty have been observed by Belle~\cite{zb} and have been interpreted in the tetraquark scheme by Ali and collaborators~\cite{Ali:2011ug}, who have also suggested to look for the $\pi^\pm J/\psi$ state in $Y(4260)$ decays.

\section*{Interpretation of the $Z_c(3900)$ state} 
The resonant mass measured by BES and Belle is close to that of the well known $X(3872)$. However, if isospin is conserved in the $Z_c^\pm $ decay, the simplest quantum numbers would be $J^{PG}=1^{++}$, with $G=G$-parity. Therefore the neutral isospin partner $Z_c^0$ would have negative charge conjugation, opposite to the $C$-conjugation of $X(3872)$ which decays into channels with  positive $C$, namely $J/\psi + \rho^0/\omega^0$ (the issue of possible isospin violation in $X$ and $Z$ decays is discussed later).  

The tetraquark model for hidden charm states discussed in Ref.~\cite{Maiani:2004vq} (see also~\cite{revs}; for earlier ideas, see~\cite{jw,exotica})  predicts {\it three} isospin multiplets with $I=1$ and $J^P=1^+$, corresponding to states of the form $[cq]_{S}[{\bar c}{\bar q}^\prime]_{S^\prime}$, with $q,q^\prime=u,d$ and spin distribution $(S,S^\prime)=(1,0), (0,1),(1,1)$. The neutral states, with $q=q^\prime=u,d$, divide into positive and negative $C$ states
\bea
&&C=+1:\; \frac{|1,0\rangle+|0,1\rangle}{\sqrt{2}};\; \; (3872~\text{MeV},\;{\rm input})\nonumber \\
&& C=-1:\;\frac{|1,0\rangle-|0,1\rangle}{\sqrt{2}};\; \; (3882~\text{MeV},\;{\rm computed} )\nonumber \\
&& C=-1: \;|1,1\rangle_{J=1};\; \; (3755~\text{MeV},\;{\rm computed} )
\label{tetraq}
\eea
Each state is made by two, almost degenerate, $I=0,1$ states. Furthermore states with same $C$ parity can also mix.

We give in parentheses the mass in MeV computed in~\cite{Maiani:2004vq} on the basis of the spin-spin interaction derived from baryon and meson spectra, extrapolated to the tetraquark under the assumption of one-gluon exchange. 
The $C=1$ state is identified with the $X(3872)$ and its mass taken as input value to fix the charm diquark mass. The mass of the higher $C=-1$ state agrees with the position of the possible resonance in the $\pi^\pm J/ \psi$ mass distribution. More precise data are needed to decide on the  existence of the second peak at a lower mass.

It is interesting to note that the prediction of the tetraquark model is different, in this case, from the molecular picture~\cite{mol}. In the latter case, one would associate $S$-wave bound states to the $D \bar D^*$ and  $ D^* \bar D^*$ thresholds. One finds also in this case three $J^P=1^+$ states, to wit
\bea
&& C=+1:\; \frac{|D,{\bar D}^*\rangle+|{\bar D}, D^*\rangle}{\sqrt{2}} ;  \; (3872~\text{MeV})   \nonumber \\ 
&& C=-1:\; \frac{|D,{\bar D}^*\rangle-|{\bar D}, D^*\rangle}{\sqrt{2}}  ;  \; (3872~\text{MeV}) \nonumber \\
&& C=-1:\; |D^*,{\bar D}^*\rangle_{J=1}  ;  \; (4014 ~\text{MeV})
\eea
where for each state we have given the value of the corresponding threshold for the neutral $D$ and $D^*$ mesons. Loosely bound charged molecules should differ by few~MeV's.
There are still two states of opposite $C$ approximately degenerate and close to the $X(3872)$ but the third state is at mass higher  than the $X(3872)$ and $Z_c(3900)$ masses. 

\begin{figure}[tb!]
 \begin{minipage}[c]{8cm}
   \centering
   \includegraphics[width=8cm]{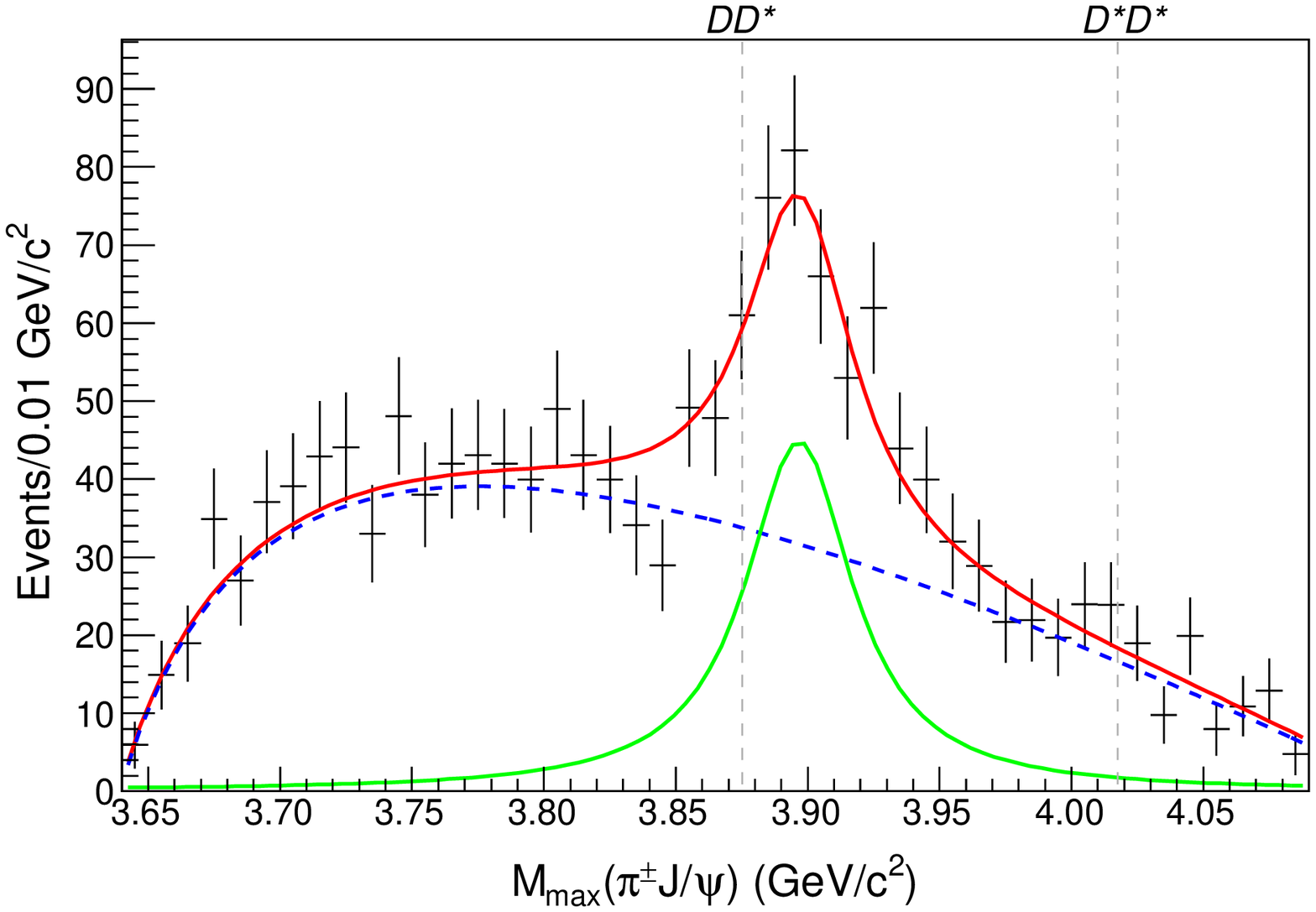}
   \includegraphics[width=8cm]{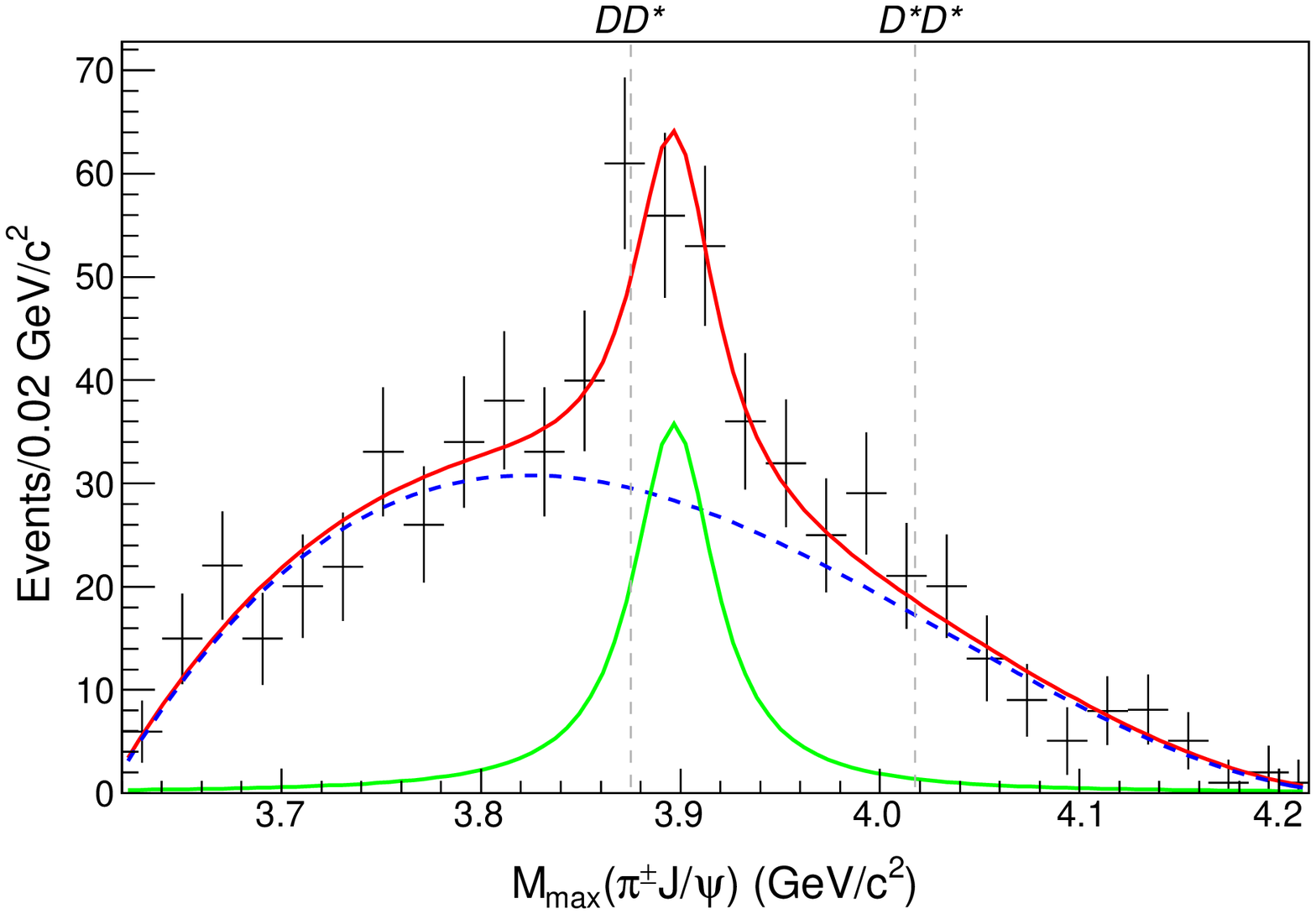}
 \end{minipage}%
 \begin{minipage}[c]{5cm}
  \centering\small
    \begin{tabular}{||c|c||} \hline
$A_\text{BES}$ & $0.96 \pm 0.04$ \\ \hline
$B_\text{BES}$ & $45 \pm 6$ \\ \hline\hline
$A_\text{Belle}$ & $1.02 \pm 0.07$ \\ \hline
$B_\text{Belle}$ & $36 \pm 7$ \\ \hline\hline
$M_1\,(\textrm{MeV})$ & $3897 \pm 3$\\ \hline
$\Gamma_1\,(\textrm{MeV})$ & $49 \pm 8$\\ \hline\hline
$\chi^2 / \textrm{DOF}$ & $50/69$  \\ \hline
CL & $96\%$ \\ \hline
\end{tabular}
 \end{minipage}
 \caption{\small Result of the combined fit to the BES~\cite{bes} (upper panel) and Belle~\cite{belle} data with one resonance only.
The thresholds for $X \to \bar D^{(*)} D^{(*)}$ are reported, where the molecular hypothesis expects the presence of resonances. The CL is the confidence level for the fit. Upper solid curves: fit to data. Lower solid curves: individual resonances. Dashed curves: background 
as in BES and Belle papers.}
\label{besplotnull}
\end{figure}

To illustrate our point, we have performed a combined fit to the $M_{\text{max}}(J/\psi\,\pi^\pm)$  distribution as published by BES~\cite{bes} and Belle~\cite{belle},  including the second resonance $Z_c^\prime$.  For this purpose a model including the same background shapes, $\text{bkg}_{\text{BES}}(x)$ and $\text{bkg}_{\text{Belle}}(x)$, as in the original papers  and two Breit-Wigner functions for the resonant states was used 
\begin{equation}
\begin{split}
 f_{\text{BES}}\left(x=M_{\text{max}}(J/\psi\,\pi^\pm) \right) &= A_{\text{BES}}\cdot\text{bkg}_{\text{BES}}(x)+B_{\text{BES}}\left| \sum_{i=1,2}  \frac{\sqrt{C_i}e^{i\phi_i} \Gamma_i M_i}{x^2-M_i^2+i \Gamma_i M_i}\right|^2\\
 f_{\text{Belle}}\left(x=M_{\text{max}}(J/\psi\,\pi^\pm) \right) &= A_{\text{Belle}}\cdot\text{bkg}_{\text{Belle}}(x)+B_{\text{Belle}} \left| \sum_{i=1,2}  \frac{\sqrt{C_i}e^{i\phi_i} \Gamma_i M_i}{x^2-M_i^2+i \Gamma_i M_i}\right|^2
 \end{split}
 \label{eq:model2}
\end{equation}
In this equations we put $C_1=1$ and $\phi_1=0$ as normalization parameters. Hence, $C_2$ is the ratio between the amplitudes. 


Assuming no significant difference between the two experiments in the ratio of the efficiencies for the two resonances, we force the two $C_2$ parameters to be the same.

\begin{figure}[htb!]
 \begin{minipage}[c]{8cm}
   \centering
   \includegraphics[width=8cm]{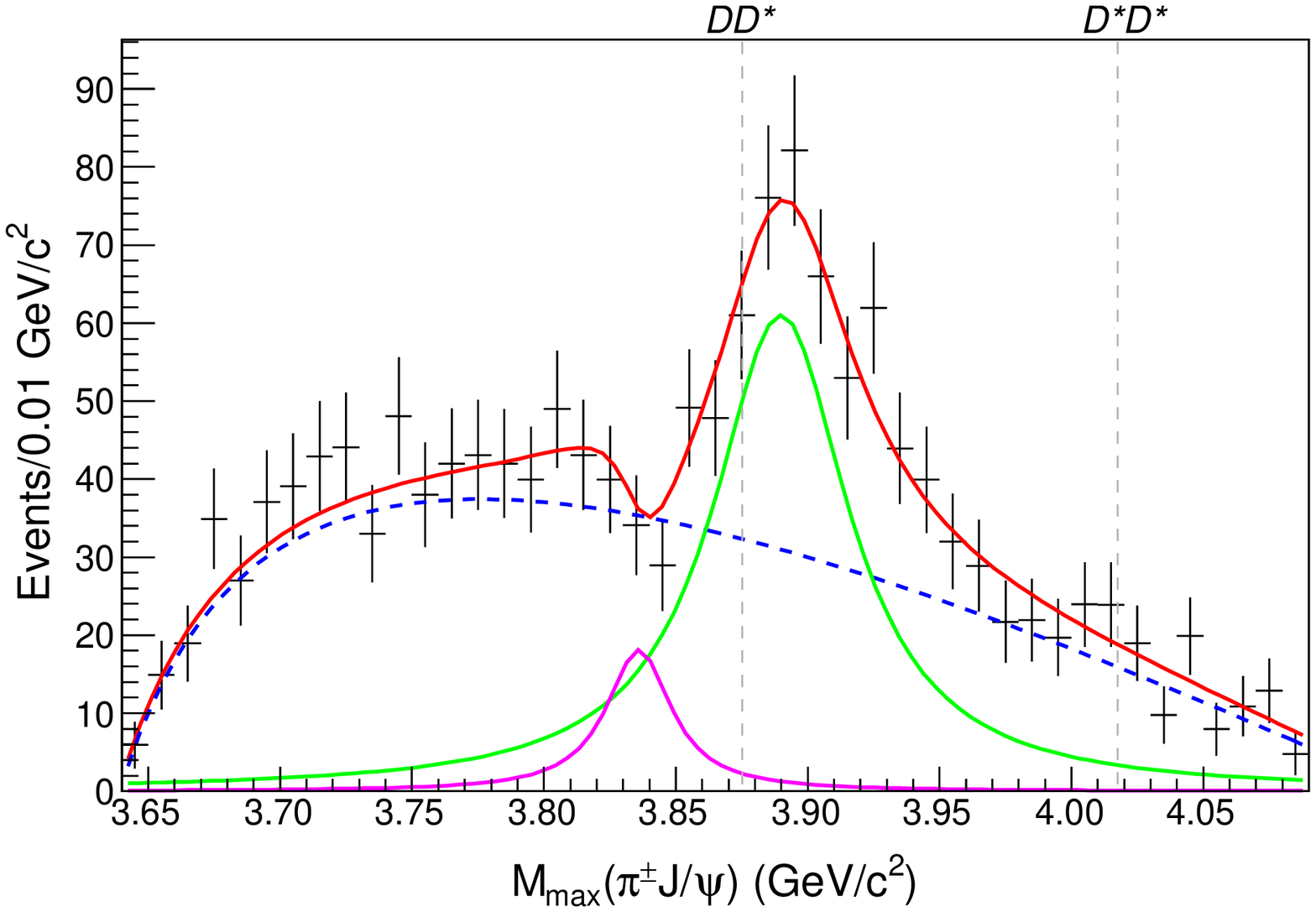}
   \includegraphics[width=8cm]{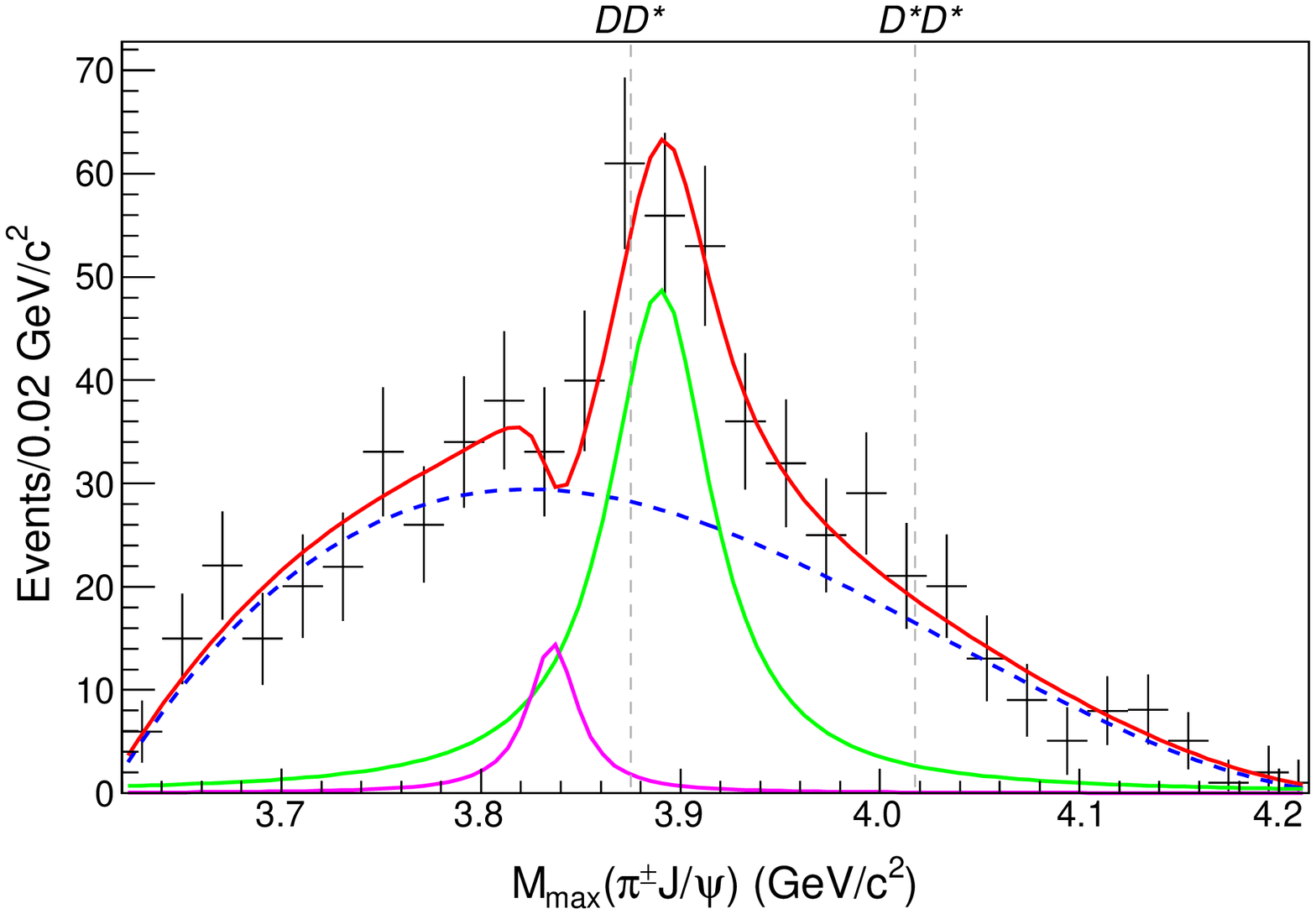}
 \end{minipage}%
 \begin{minipage}[c]{5cm}
  \centering\small
    \begin{tabular}{||c|c||} \hline
$A_\text{BES}$ & $0.92 \pm 0.07$ \\ \hline
$B_\text{BES}$ & $61 \pm 18$ \\ \hline\hline
$A_\text{Belle}$ & $0.98 \pm 0.09$ \\ \hline
$B_\text{Belle}$ & $49 \pm 15$ \\ \hline\hline
$M_1\,(\textrm{MeV})$ & $3890 \pm 6$\\ \hline
$\Gamma_1\,(\textrm{MeV})$ & $63 \pm 12$\\ \hline\hline
$M_2\,(\textrm{MeV})$ & $3836 \pm 13$\\ \hline
$\Gamma_2\,(\textrm{MeV})$ & $30 \pm 18$\\ \hline
$C_2$ & $0.30 \pm 0.25$ \\ \hline
$\phi_2\,(\textrm{DEG})$ & $109 \pm 33$\\ \hline\hline
$\chi^2 / \textrm{DOF}$ & $41/65$  \\ \hline
CL & $99\%$ \\ \hline
\end{tabular}
 \end{minipage}
 \caption{\small Result of the combined fit to the BES~\cite{bes} (upper panel) and Belle~\cite{belle} data with the model described in the text.
The thresholds for $X \to \bar D^{(*)} D^{(*)}$ are reported, where the molecular hypothesis expects the presence of resonances. The CL is the confidence level for the fit. Upper solid curves: fit to data. Lower solid curves: individual resonances. Dashed curves: background 
as in BES and Belle papers.}
\label{besplot}
\end{figure}

As a start we performed the fit assuming only one state (see~Fig.\ref{besplotnull}), i.e. forcing $C_2=0$. 
The estimated mass and width are consistent with those published by BES and the $\chi^2/\text{DOF}=50/69$, with a CL $=96\%$.  It is interesting to note that such a good $\chi^2$ stresses the consistency of the resonant structure in the  two datasets.  

\begin{figure}[htb!]
 \begin{minipage}[c]{8cm}
   \centering
   \includegraphics[width=8cm]{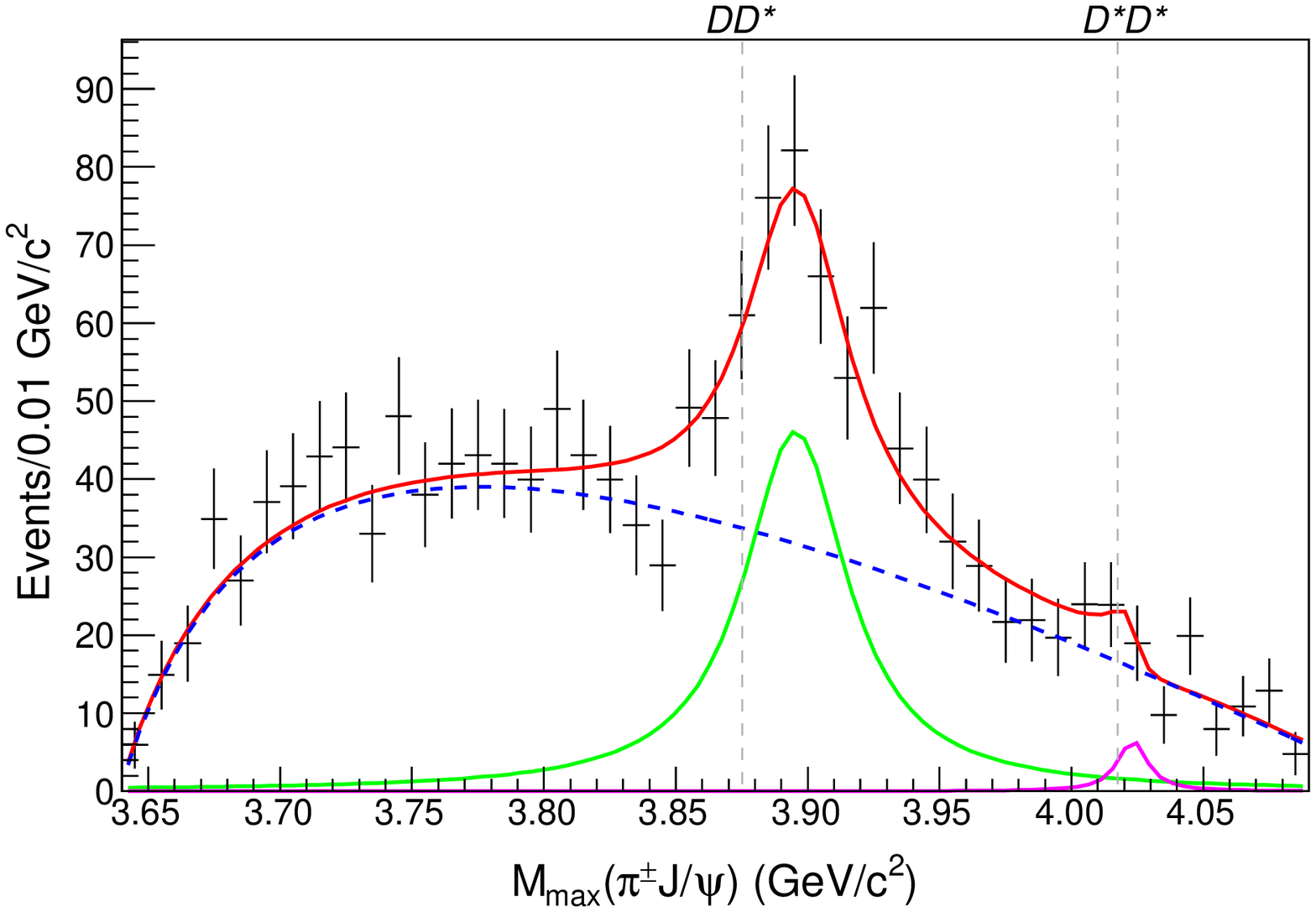}
   \includegraphics[width=8cm]{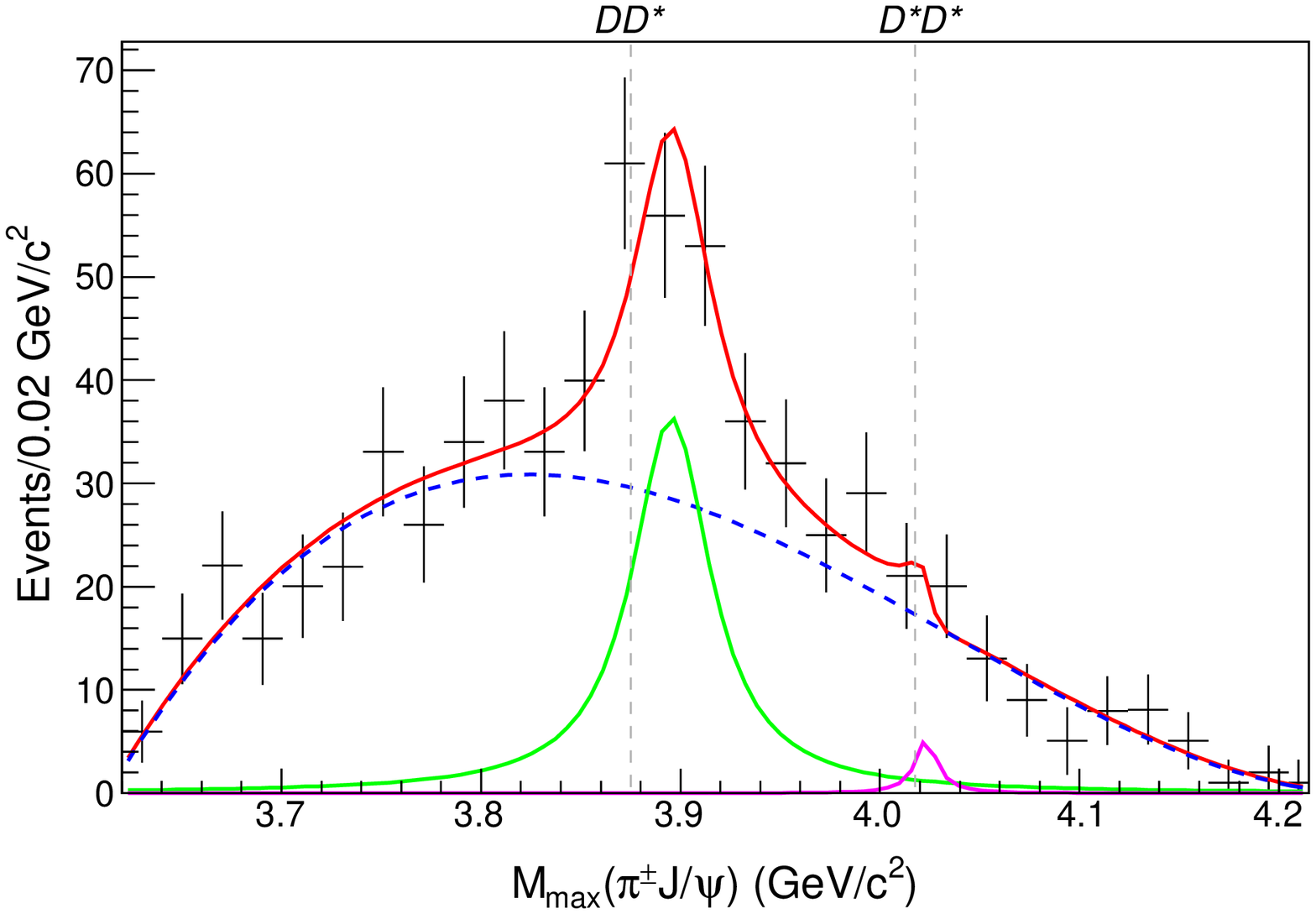}
 \end{minipage}%
 \begin{minipage}[c]{5cm}
  \centering\small
  \begin{tabular}{||c|c||} \hline
$A_\text{BES}$ & $0.96 \pm 0.05$ \\ \hline
$B_\text{BES}$ & $46 \pm 6$ \\ \hline\hline
$A_\text{Belle}$ & $1.03 \pm 0.06$ \\ \hline
$B_\text{Belle}$ & $36 \pm 7$ \\ \hline\hline
$M_1\,(\textrm{MeV})$ & $3895 \pm 3$\\ \hline
$\Gamma_1\,(\textrm{MeV})$ & $48 \pm 8$\\ \hline\hline
$M_2\,(\textrm{MeV})$ & $4023 \pm 6$\\ \hline
$\Gamma_2\,(\textrm{MeV})$ & $13 \pm 26$\\ \hline
$C_2$ & $0.14 \pm 0.31$ \\ \hline
$\phi_2\,(\textrm{DEG})$ & $196 \pm 77$\\ \hline\hline
$\chi^2 / \textrm{DOF}$ & $47/65$  \\ \hline
CL & $95\%$ \\ \hline\end{tabular}
 \end{minipage}
 \caption{\small Result of the combined fit to the BES~\cite{bes} (upper panel) and Belle~\cite{belle} data with the model described in the text but forcing the second resonance to be above the $Z_c(3900)$ peak. 
The thresholds for $X \to \bar D^{(*)} D^{(*)}$ are reported, where the molecular hypothesis expects the presence of resonances. The CL is the confidence level for the fit. Upper solid curves: fit to data. Lower solid curves: individual resonances. Dashed curves: background 
as in BES and Belle papers.}
\label{besplotmol}
\end{figure}

When we allow for the second resonance (Fig.~\ref{besplot}), its fitted mass is  \mbox{$54\pm 14$~MeV} below the main resonance and the fit yields $\chi^2/\text{DOF}=41/65$. A negative interference is also evidenced.
Finally, in order to test the molecular hypothesis, we have repeated the fit forcing the mass of the second structure to be above the $Z_c(3900)$. In this case (see Fig.~\ref{besplotmol}) the $\chi^2$ increases to 47 with 65 DOF and there is no hint for such a state.  

In order to quantify the significance of the second structure included in the fits, we have adopted the statistical approach described in detail in Ref.~\cite{Faccini:2012zv}: from the fit to the data performed assuming only one exotic structure, we have simulated a large number of mock experiments, correctly accounting for statistical fluctuations. On each of them we have performed three fits, one assuming only one exotic resonance, the other one assuming a second resonance with a mass smaller than the dominant one (the ``tetraquark'' assumption),
and the last one assuming a second resonance with a mass larger than the dominant one (the ``molecular'' assumption). For each mock experiment we have recorded the $\chi^2$ of each fit, called $\chi^2_{0}$, $\chi^2_\text{tetra}$ and $\chi^2_\text{mol}$, respectively. From the distribution of $\Delta\chi^2_\text{tetra}=\chi^2_\text{tetra}-\chi^2_0$, we can estimate the probability of a second structure to appear before the main one in absence of a real signal as the fraction of mock experiments where $\Delta\chi^2_\text{tetra}<\Delta\chi^2_\text{tetra;data}=41-50=-9$. In this way we have estimated that there is only a $12\%$ probability of the second structure as fitted in Fig.~\ref{besplot} to be a statistical fluctuation. The same procedure applied to the ``molecular'' assumption, where $\Delta\chi^2_\text{mol;data}=47-50=-3$, yields a $53\%$ probability of the structure being due to a statistical fluctuation.


\section*{Isospin breaking}
Isospin symmetry in QCD is broken by the $u-d$ quark mass difference and by second-order photon and weak boson exchange processes. 

We introduce isospin breaking by inserting the corresponding effective lagrangian in the isospin symmetric amplitudes. This gives a small effect, of the order of few percent, {\it except} for the cases where the effective lagrangian is inserted in the external legs. In the the latter case, if there are near degeneracies between states of different isospin, the effect is enhanced by the presence of small denominators. The situation occurs for the neutral members of the $I=0$ and 1 multiplet considered before, where the mass difference may be of the same order of the $u-d$ quark mass difference, as it was argued in~\cite{Maiani:2004vq}. 

The upshot of this discussion~\cite{Maiani:2008zz}
is that when considering the charged tetraquarks we may assume isospin and $G$-parity as good quantum numbers, to a few percent accuracy, and use them to derive selection rules for production and decay. Decays of neutral states are described by isospin conserving amplitudes from states with definite isospin but we have to use initial states with definite superpositions of $I=0,1$. The only good quantum number, in this case, is $C$-conjugation.

Finally, it should be noted that in some molecular descriptions, isospin breaking is related to the distance between thresholds~\cite{Hanhart:2011tn}. For example, the mass of $X(3872)$ is just at $D^0 \bar D^{*0}$ threshold, but is $\sim 8$~MeV below the $D^+ D^{*-}$ threshold. Being $\Gamma_X \ll 8$~MeV, a molecular $X$ cannot have a big $D^+ D^{*-}$ component, thus it cannot be a pure $I=0$ state. On the other hand, the mass of $Z_c(3900)$ is above both thresholds, and $\Gamma_{Z_c} \gg 8$~MeV, thus it should be a pure $I=1$ state, and no isospin-violating decay should occur.

\section*{Decay widths}

While the observed mass in Fig.~\ref{besplot} is remarkably close to the predicted $3882$~MeV value, the width is significantly larger than that of the $X(3872)$ ($\Gamma(3872)\lesssim 2$~MeV). 

To explain this, we considered 
that the  $J/\psi\, \pi$ and $\psi(2S) \pi$ channels have a definitely larger phase space than the  $J/\psi \,\rho,\omega$. 
The slightly higher mass of $Z_c^+$ allows, in addition,  for larger decay rates into $D\bar D^*$.  

For a crude estimate of the decay width of $Z_c^+ \to D \bar D^*$, we follow Ref.~\cite{brazzi} (see Table~II therein). We start from the expression of the decay rate:
\begin{multline}
 \Gamma\left(X(3872)\to D^0 \bar D^{*0} + \mathrm{c.c.} \right) =\\= \frac{p^*\left(M_X,M_{D^0},M_{D^{*0}}\right)}{8\pi M_X^2} \frac13 g^2_{XDD^*} \left(3 + \frac{p^*\left(M_X,M_{D^0},M_{D^{*0}}\right)^2}{M_{D^{*0}}^2}\right)\notag
\end{multline}
where $g_{XDD^*}$ is a coupling with the dimension of a mass. The decay is phase space forbidden, so we average over a random mass of the $X$ extracted from a Breit-Wigner centered at $3871.68$~MeV with $\Gamma(3872)=1.2$~MeV (the experimental resolution) and $M_{D^0} + M_{D^{*0}} < M_X < M_B - M_K$. 
We get:
\begin{equation}
 g_{XDD^*} = 2.5\textrm{~GeV}\notag
\end{equation}

Next, we assume $g_{Z_c^+ D\bar D^*}\simeq g_{XD\bar D^*}$ and obtain:
\be
\Gamma(Z_c^+ \to D^+ \bar D^{*0}, \bar D^0 D^{*+})\approx 4~ {\rm MeV}
\ee

For the other decay modes we offer the following estimate:
\begin{enumerate}
 \item $\Gamma(Z_c^+ \to J/\psi \;\pi^+)\approx 29$ MeV
 \item $\Gamma(Z_c^+ \to \psi(2S) \;\pi^+)\approx 6$ MeV
 \item $\Gamma(Z_c^+ \to \eta_c \;\rho^+)\approx 19$ MeV
\end{enumerate}
 We rely on a rough dimensional argument adopting $g \approx M_{Z_c^+}\approx 3.9\textrm{~GeV}$ for the unknown couplings. All in all, we get to a total width of $60$~MeV. 

As for the $J^{PC}=1^{+-}$ neutral state we have four $D \bar D^{*}$ decay modes, $D^0 \bar D^{*0}$, etc., which give a width of $7$~MeV. Including the decay into $J/\psi\; \eta$ (assuming maximal isospin breaking) and the same decays as for the charged component we estimate a total width $\Gamma(Z_c^0)\approx 80$~MeV.

In the case of $(Z_c^\prime)^+$ we will use $g \approx M_{(Z_c^\prime)^+}\approx 3.8\textrm{~GeV}$ for the unknown couplings. The charged state widths can be estimated as:
\begin{enumerate}
 \item $\Gamma((Z_c^\prime)^+ \to J/\psi \;\pi^+)\approx 24$~MeV
 \item $\Gamma((Z_c^\prime)^+ \to \eta_c \;\rho^+)\approx 6$~MeV
\end{enumerate}
Accounting for a total width of $\Gamma((Z_c^\prime)^+)\approx 30$~MeV.

The $Z_c^\prime$ neutral state also has the isospin violating decay $(Z_c^\prime)^0 \to J/\psi\; \eta$ for which we get $\Gamma \approx 12$~MeV.
Therefore for the neutral state we can estimate a total width of $\Gamma((Z_c^\prime)^0)\approx40$~MeV. 

Given our ignorance on the couplings, these results have to be taken as mere order of magnitude estimates.

Nevertheless it should be remarked that the assumption $g_{Z_c^+ D\bar D^*}\simeq g_{XD\bar D^*}$ leads to a $\mathcal{B}\left(Z_c^+ \to D\bar D^*\right)\sim 10\%$. On the countrary, in the molecular picture, the decay of the molecule into its open charm constituents should be dominant over the short-range decays into charmonium and light mesons~\cite{Voloshin:2013dpa,Braaten:2005ai}, for example, in the case of the X,  it is known that $\mathcal{B}\left(X(3872)\to D\bar D^*\right)\sim70\%$ and this is considered as one of the hints in favor of its molecular description. 

\section*{Conclusions}
The $Z_c(3900)$, just discovered by BES and Belle, well fits a state predicted by some of us within the tetraquark model. In this paper we 
reanalized the experimental spectrum testing the hypothesis of the existence of an additional resonance.
 Within the tetraquark model indeed another structure is required to occur at about 100~MeV below that peak whereas the molecule picture would require a  $D\bar D^*$ structure higher by $\sim$100~MeV. We show that there may be hints of an additional structure consistently with the expectations of the tetraquark model. 
 
Furthermore, we investigate the decay modes of the  $Z_c(3900)$ and indicate reasons for it to have a larger width than the $X(3872)$ as observed in data. From this study we also derive that a signal should be sought in  $Y(4260)\to \pi^0\pi^0 \,J/\psi$ or $Y(4260)\to \pi^0\eta \,J/\psi$ decays to search  for the neutral component of the isotriplet, due to decay in $J/\psi\,\pi^0$ and possibly  in $J/\psi\,\eta$ in presence of sizeable isospin violation.


\begin{thebibliography}{9}
  \bibitem{revs} 
  N.~Drenska, R.~Faccini, F.~Piccinini, A.~Polosa, F.~Renga and C.~Sabelli,
  Riv.\ Nuovo Cim.\  {\bf 033}, 633 (2010)
  [\href{http://arxiv.org/abs/1006.2741}{arXiv:1006.2741 [hep-ph]}];
   R.~Faccini, A.~Pilloni and A.~D.~Polosa,
  Mod.\ Phys.\ Lett.\ A {\bf 27}, 1230025 (2012)
  [\href{http://arxiv.org/abs/1209.0107}{arXiv:1209.0107 [hep-ph]}].

\bibitem{z4430} 
  S.~K.~Choi {\it et al.}  [BELLE Collaboration],
  Phys.\ Rev.\ Lett.\  {\bf 100}, 142001 (2008)
  [\href{http://arxiv.org/abs/0708.1790}{arXiv:0708.1790 [hep-ex]}];
  B.~Aubert {\it et al.}  [BaBar Collaboration],
  Phys.\ Rev.\ D {\bf 79}, 112001 (2009)
  [\href{http://arxiv.org/abs/0811.0564}{arXiv:0811.0564 [hep-ex]}].
  
\bibitem{zb} 
  A.~Bondar {\it et al.}  [Belle Collaboration],
  Phys.\ Rev.\ Lett.\  {\bf 108}, 122001 (2012)
  [\href{http://arxiv.org/abs/1110.2251}{arXiv:1110.2251 [hep-ex]}].

\bibitem{bes} 
  M.~Ablikim {\it et al.}  [BESIII Collaboration],
\href{http://arxiv.org/abs/1303.5949}{arXiv:1303.5949 [hep-ex]}.

\bibitem{belle} 
  Z.~Q.~Liu {\it et al.}  [Belle Collaboration],
  \href{http://arxiv.org/abs/1304.0121}{arXiv:1304.0121 [hep-ex]}.

\bibitem{cleo} 
T.~Xiao, S.~Dobbs, A.~Tomaradze and K.~K.~Seth,
  \href{http://arxiv.org/abs/1304.3036}{arXiv:1304.3036 [hep-ex]}.
\bibitem{Maiani:2004vq} 
  L.~Maiani, F.~Piccinini, A.~D.~Polosa and V.~Riquer,
  Phys.\ Rev.\ D {\bf 71}, 014028 (2005)
  [\href{http://arxiv.org/abs/hep-ph/0412098}{hep-ph/0412098}].
  
  \bibitem{mol} F. E. Close and P. R. Page, Phys. Lett. B {\bf 628}, 215 (2005); E. Braaten and M. Kusunoki, Phys. Rev. D {\bf 69}, 074005 (2004); F. E. Close and P. R. Page, Phys. Lett. B {\bf 578}, 119 (2004); N.A. Tornqvist, Phys. Lett. B {\bf 590}, 209 (2004); E. S. Swanson, Phys. Rep. {\bf 429}, 243 (2006); S. Fleming, M. Kusunoki, T. Mehen, and U. van Kolck, Phys. Rev. D {\bf 76}, 034006 (2007); E. Braaten and M. Lu, Phys. Rev. D {\bf 76}, 094028 (2007); D {\bf 77}, 014029 (2008).
      
    \bibitem{Ali:2011ug}
  A.~Ali, C.~Hambrock and W.~Wang,
  Phys.\ Rev.\ D {\bf 85} (2012) 054011
  [\href{http://arxiv.org/abs/1110.1333}{arXiv:1110.1333 [hep-ph]}].
\bibitem{jw} 
  R.~L.~Jaffe and F.~Wilczek,
  Phys.\ Rev.\ Lett.\  {\bf 91}, 232003 (2003)
  [\href{http://arxiv.org/abs/hep-ph/0307341}{hep-ph/0307341}].
 \bibitem{exotica} 
  R.~L.~Jaffe,
  Phys.\ Rept.\  {\bf 409}, 1 (2005)
  [\href{http://arxiv.org/abs/hep-ph/0409065}{hep-ph/0409065}].
  
  \bibitem{Faccini:2012zv} 
  R.~Faccini, F.~Piccinini, A.~Pilloni and A.~D.~Polosa,
  Phys.\ Rev.\ D {\bf 86}, 054012 (2012)
  [arXiv:1204.1223 [hep-ph]].

   \bibitem{Maiani:2008zz}
  L.~Maiani, A.~D.~Polosa and V.~Riquer,
  New J.\ Phys.\  {\bf 10} (2008) 073004.

  \bibitem{Hanhart:2011tn} 
  C.~Hanhart, Y.~.S.~Kalashnikova, A.~E.~Kudryavtsev and A.~V.~Nefediev,
  Phys.\ Rev.\ D {\bf 85}, 011501 (2012)
  [\href{http://arxiv.org/abs/1111.6241}{arXiv:1111.6241 [hep-ph]}].
 
 \bibitem{brazzi} 
  F.~Brazzi, B.~Grinstein, F.~Piccinini, A.~D.~Polosa and C.~Sabelli,
  Phys.\ Rev.\ D {\bf 84}, 014003 (2011)
  [\href{http://arxiv.org/abs/1103.3155}{arXiv:1103.3155 [hep-ph]}].
  
\bibitem{Voloshin:2013dpa} 
  M.~B.~Voloshin,
  \href{http://arxiv.org/abs/1304.0380}{arXiv:1304.0380 [hep-ph]}.
  
\bibitem{Braaten:2005ai} 
  E.~Braaten and M.~Kusunoki,
  Phys.\ Rev.\ D {\bf 72}, 054022 (2005)
  [\href{http://arxiv.org/abs/hep-ph/0507163}{hep-ph/0507163}].

\end{thebibliography}
\end{document}